\documentstyle[12pt,dina4p]{article}

\begin{document}
\begin{flushright}
hep-th/9512056
\end{flushright}   \vspace{10mm}
\begin{center}
{{\Large \bf
Non-Commutative Geometry, Multiscalars,\\ and the
Symbol Map
}}
\footnote{To appear in the Proceedings of the 29.
          International Symposium on the Theory of Elementary Particles,
          Buckow, 1995.   }
  \\
\vspace{5mm}
{\sc M.\, Reuter}\\
{\it Deutsches Elektronen-Synchrotron DESY,\\
     Notkestrasse 85, D-22603 Hamburg, Germany}
\end{center}
\vspace{1mm}
\begin{abstract}
Starting from the concept of the universal
exterior algebra in non-commutative differential geometry
we construct differential forms on the quantum phase-space of an
arbitrary system. They bear the same natural relationship to
quantum dynamics which ordinary tensor fields have with respect
to classical hamiltonian dynamics.
\end{abstract}
\renewcommand{\theequation}{1.\arabic{equation}}
\setcounter{equation}{0}

 \section{Introduction}
The mathematical setting of the standard hamiltonian formalism
is the classical geometry of symplectic manifolds.
Therefore all concepts and constructions of classical differential
geometry (vectors, forms, exterior and Lie derivatives, \dots)
can be applied to the study of phase-spaces and the
dynamics on them.
After quantization most of these notions are lost because the
position and momentum variables, previously coordinates of
phase-space, are non-commuting operators then.
In the following we address the question of what happens to
these geometrical constructions at the quantum
level. In particular, we shall present a construction of the
quantum analogue of the exterior algebra. It combines
the universal differential forms which appear in the
non-commutative geometries of A.Connes \cite{co}
with the (Wigner--Weyl) symbol map\cite{lich}.
The main virtue of this idea is that quantum and classical tensor fields
are represented in a unified framework now, the former being
a smooth deformation of the latter \cite{mr}.
We shall see that the quantum differential forms are represented
by multiscalar functions which depend on more than one
phase-space argument. As a warm-up, we first discuss classical
forms in a similar framework.

Let $\bigwedge^p_{\rm MS}$ denote the set of ``multiscalar"
functions $F_p (\phi_0, \phi_1, \cdots, \phi_p)$ which
depend on $p + 1$ arguments $\phi_i, i = 0, 1, \cdots,
p$ and which vanish if two neighboring arguments are
equal \cite{review}.
Now we define a map
$
\delta :  \bigwedge \nolimits^p_{\rm MS} \rightarrow
\bigwedge \nolimits^{p+1}_{\rm MS}
$
by
\begin{equation}
\label{2.2}
\fbox{$ \displaystyle
\left ( \delta F_p \right ) \left ( \phi_0, \cdots, \phi_{p+1} \right )
= \sum_{i = 0}^{p+1} (-1)^i \, F_p \left ( \phi_0, \cdots,
\phi_{i-1}, \widehat{\phi}_i, \phi_{i+1}, \cdots, \phi_{p+1} \right)
$}
\end{equation}
where the caret over $\phi_i$ means that this argument
is omitted. The $\delta$--operation maps a function of
$p+1$ arguments onto a function of $p+2$ arguments.
Remarkably enough, $\delta$ turns out to be nilpotent:
$
\delta^2 = 0.
$
Henceforth we shall refer to a function $F_p \in \bigwedge^p_{\rm MS} $1
as a ``$p$--form''. On the direct sum
$
\bigwedge \nolimits^\ast_{\rm MS} \, = \, \bigoplus^\infty_{p = 0} \,
\bigwedge \nolimits^p_{\rm MS}
$
there exists a natural product of a $p$--form $F_p$ with
a $q$--form $G_q$ yielding a $(p + q)$--form $F_p \bullet G_q$:
\begin{equation}
\left( F_p \bullet G_q \right) \left(\phi_0,
\cdots, \phi_p, \phi_{p+1}, \cdots, \phi_{p+q} \right)
= F_p \left(\phi_0, \phi_1, \cdots, \phi_p \right) G_q \left(\phi_p,
\phi_{p+1}, \cdots, \phi_{p+q} \right)
\end{equation}
With respect to this product
$\delta$  obeys the Leibniz rule
$
\delta \left(F_0 \bullet G_0 \right) = \left(\delta F_0 \right)
\bullet G_0 + F_0 \bullet \left(\delta G_0 \right).
$

Now
we assume that the $\phi_i \equiv \left(\phi_i^a \right)$
are local coordinates on some manifold ${\cal M}$, and that the
$F_p$'s are smooth functions which transform as multiscalars
under general coordinate transformations (diffeomorphisms)
on ${\cal M}$. This means in particular that $F_p$ evolves
under the flow generated by some vector field
$h = h^a (\phi) \partial_a, \partial_a \equiv \frac
{\partial}{\partial \phi^a}$,
according to
$
\partial_t  F_p  \left(t \right)
= \sum_{i = 0}^{p} {\rm V}  \left( \phi_i \right)
F_p \left(t \right)
$,
where
$
{\rm V}  \left(\phi_i \right) = - h^a \left(\phi_i \right)
\partial_a^{(i)}
$
acts only on the $\phi_i$-argument of $F_p$. Here the
``time'' $t$ parametrizes points along the flow lines
of the vector field $h$.
In the following we restrict our attention to symplectic manifolds
${\cal M} \equiv {\cal M}_{2N}$ and to
vector fields which are hamiltonian, i.e. we assume that
(locally)
$
h^a (\phi) = \omega^{ab} \partial_b H (\phi)
$
for some generating function $H$.
Thus the Lie derivative, i.e., (minus) the RHS of the evolution
equation, becomes
$
{\cal L}_p [H] = - \sum_{i = 0}^{p}
\omega^{ab}
\partial_a H (\phi_i)
 \partial^{(i)}_b
$.

Next we show how in the limit when the
arguments of the multiscalar $F_p$
are very ``close'' to each other, the generalized
$p$--form $F_p \in \bigwedge^p_{\rm MS}$ gives rise to
a conventional $p$--form. We set
$
\phi^a_0  =  \phi^a, \
\phi^a_i  =  \phi^a  + \eta^a_i \, , \, i = 1, \cdots, p
$
and expand $
F_p \left( \phi, \phi + \eta_1, \cdots, \phi + \eta_p \right)$~
to lowest order in $\eta_i^a$.
We keep only terms in which all $\eta_i$'s are different
and obtain a sum of terms of the type
\begin{equation}
\eta^{a_1}_{i_1} \eta^{a_2}_{i_2} \cdots \eta^{a_l}_{i_l}
 \, \, \partial^{(i_1)}_{a_1} \cdots \partial^{(i_l)}_{a_l}
 F_p (\phi, \cdots, \phi)
\end{equation}
for $0 \le l \le p$.
After ``stripping off'' the $\eta$'s, the quantities
$\partial^{(i_1)}_{a_1} \cdots \partial^{(i_l)}_{a_l}
F_p (\phi, \phi, \cdots, \phi)$, for $i_1, \cdots, i_l$ fixed,
transform as the components of a covariant tensor field of rank
$l$, because on each $\phi$--argument there acts at most one
derivative. Upon explicit antisymmetrization in
the indices $a_1, \cdots, a_l$ we obtain the components of
an $l$--form.
Because the functions in $
\wedge^p_{\rm MS}$ vanish if two neighboring arguments are equal
one finds that the
expansion contains only the term with the
maximal rank $l = p$:
\begin{equation}
\label{2.20}
F_p  \left( \phi, \phi + \eta_1, \cdots, \phi + \eta_p \right)
 = \eta^{a_1}_1 \eta^{a_2}_2 \cdots  \eta^{a_p}_p
\, \, \partial^{(1)}_{a_1} \cdots \partial^{(p)}_{a_p}
F_p (\phi, \phi, \cdots, \phi) + O \left(\eta^2_i \right)
                          \end{equation}
It is convenient to look at the ordinary differential
forms induced by the multiscalars
 as the image of the so-called
``classical map'' \cite{review}
$
{\rm Cl}: \bigwedge \nolimits ^p_{\rm MS} \left({\cal M}_{2N}\right)
\rightarrow \bigwedge \nolimits ^p \left({\cal M}_{2N} \right)
$
which is defined by
\begin{equation}
\left[ {\rm Cl} \left( F_p \right) \right]
(\phi) = \partial^{(1)}_{a_1} \cdots
 \partial^{(p)}_{a_p} F (\phi, \cdots, \phi)
\, d \phi^{a_1} \wedge \cdots \wedge d \phi^{a_p}
\end{equation}
Under the classical map the various operations defined for multiscalars
are mapped onto their counterparts for ordinary differential forms:
\begin{equation}
{\rm Cl} \left( F_p \bullet G_q \right) = {\rm Cl} (F_p)
\wedge {\rm Cl} (G_q)    ,\
{\rm Cl} \left( \delta F_p \right) = d \, {\rm Cl} (F_p)     ,\
{\rm Cl} \left ( {\cal L}_p [H] F_p \right) = l_h
{\rm Cl} (F_p)
\end{equation}

\renewcommand{\theequation}{2.\arabic{equation}}
\setcounter{equation}{0}

\section{Universal Differential Forms}
Let us briefly review some properties of the
universal differential forms in non--commutative
geometry \cite{co,review}.
To any algebra $A$ we can associate its universal differential
envelope $\Omega A$, the algebra of ``universal differential
forms''.
To each element $a \in A$
we associate a new object $\delta a$. As a vector space,
$\Omega A$ is defined to be the linear space of words
built from the symbols $a_i \in A$ and $\delta a_i$, e.g.,
$a_1 \delta a_2 a_3 a_4 \delta a_3$. The multiplication
in $\Omega A$ is defined to be associative and distributive
over the addition $+$. The product of two elementary words
is obtained by simply concatenating the two factors.
One imposes the Leibniz rule
$
\delta \left( a_1 a_2 \right) = \left( \delta a_1 \right)
a_2 + a_1 \delta a_2
$.
By virtue of this relation, any element of $\Omega A$ can be
rewritten as a sum of monomials of the form
$
a_0 \delta a_1 \delta a_2 \cdots \delta a_p$~or
$   \delta a_1 \delta a_2 \cdots \delta a_p           $.
In order to put the two types of monomials
on an equal footing it is convenient to add a
new unit ``1'' to $A$, which is different from the unit
$A$ might have had already. We require $\delta 1 = 0$.
As a consequence, we have to consider only words of the first
type because the second one obtains for
$a_0 = 1$ then.
Finally one defines a linear operator $\delta$ by the rules
$
\delta^2 = 0$~and $
\delta \left( a_0 \delta a_1 \delta a_2 \cdots \delta a_p \right)
 =  \delta a_0 \delta a_1 \delta a_2 \cdots \delta a_p
$.
By definition, $\Omega^p A$ is the linear span of
the words $a_0 \delta a_1 \cdots \delta a_p$, referred to
as ``universal $p$--forms''. Then
\begin{equation}
\Omega A = \bigoplus^\infty_{p = 0}
\Omega^p A    ,    \ \ \ \ \   \Omega^0 A \equiv A,
\end{equation}
is a graded differential algebra with the ``exterior derivative''
\begin{equation}
\delta: \Omega^p A \longrightarrow \Omega^{p+1} A
\end{equation}

The space
$\Omega^p A$ can be identified
with a certain subspace of the tensor product
$
A \otimes A \otimes \cdots \otimes A
\equiv A^{\otimes (p+1)}
$.
Let us start with a few definitions. The associative
$\otimes_A$--product of
elements from $A^{\otimes (p+1)}$ with elements from
$A^{\otimes (q+1)}$ yields elements in
$A^{\otimes (p + q + 1)}$. It is defined by
\begin{equation}
\left[ a_0 \otimes             \cdots \otimes a_p \right]
 \otimes_A
\left[ b_0 \otimes  \cdots \otimes b_q \right]
= a_0 \otimes  \cdots \otimes a_{p-1}
\otimes a_p b_0 \otimes b_1 \otimes \cdots \otimes b_q
\end{equation}
where $a_p b_0$ is an ordinary algebra product.
It is also convenient to introduce the linear multiplication maps
${\rm m}_i
$ according to              \begin{equation}
{\rm m}_i \left[ a_0 \otimes  \cdots \otimes a_{i-1}
\otimes a_i \otimes \cdots \otimes a_p \right]
= a_0 \otimes \cdots \otimes a_{i-1} a_i
\otimes a_{i+1} \otimes \cdots \otimes a_p.
\end{equation}
Then the construction of $\Omega A$ is as follows.
We set $\Omega^0 A \equiv A$, and we identify
$ \delta a \in \Omega^1 A$ with
$
\delta a = 1 \otimes a - a \otimes 1  \in
A \otimes A
$.
``Words'' are formed by taking $\otimes_A$--products of $a$'s
and $\delta a$'s.
A generic element of $\Omega^1 A$ has the structure
$
a \delta b
= a \otimes b - ab \otimes 1
$
and is in the kernel of
${\rm m}_1$~therefore:
${\rm m}_1 (a \delta b) = ab - ab = 0$. More generally one defines
\begin{equation}
\Omega^p A\  =\ \Omega^1 A \otimes_A \Omega^1 A
\otimes_A \cdots \otimes_A \Omega^1 A
\end{equation}
where the product consists of $p$~factors.

Let us assume that it is possible to enumerate the
elements of $A$ as $ A= \{ a_m, m \in \Im \}$ where
$\Im$ is some index set.
Then a generic $p$--form $\alpha_p
\in \Omega^p A$ has an expansion
\begin{equation}
\label{3.16}
\alpha_p = \sum_{m_0 \cdots m_p}  \, \alpha_{m_0 \cdots
m_p} \, \, a_{m_0} \otimes a_{m_1} \otimes \cdots \otimes
a_{m_p}
\end{equation}
in which the coefficients
 $\alpha_{m_0 \cdots m_p}$~are
 subject to the constraints which follow from
 $
 {\rm m}_i \alpha_p = 0$. In this language, the action of the map
$\delta$~is given by \cite{mr}
\begin{equation}
\fbox{$ \displaystyle
\delta \alpha_p = \sum^{p+1}_{i=0} (- 1)^i
\sum_{m_0 \cdots m_p} \alpha_{m_0 \cdots m_p}
\, \, a_{m_0} \otimes a_{m_1} \otimes \cdots
 \otimes a_{m_{i-1}} \otimes 1 \otimes a_{m_i} \otimes
 \cdots \otimes a_{m_p}
$}
 \end{equation}
\renewcommand{\theequation}{3.\arabic{equation}}
\setcounter{equation}{0}

 \section{Quantum Forms on Phase--Space}

Let us recall some elements
of the phase--space formulation of quantum mechanics
\cite{lich}. We consider a set of operators
$\hat{f}, \hat{g}, \cdots$ on some Hilbert space ${\cal H}$, and
we set up a one--to--one correspondence between these operators
and the complex--valued functions $f, g, \cdots \in
{\rm Fun}({\cal M})$ defined over a suitable manifold ${\cal M}$.
                                                           We write
$ f = {\rm symb}~(\hat{f})$,
                       and we refer to the function $f$ as the
symbol of the operator $\hat{f}$. The symbol map
``symb'' is linear and has a well--defined inverse. An
important notion is the ``star product'' which implements the
operator multiplication at the level of symbols:
$
{\rm symb}(\hat{f} \, \hat{g}) = {\rm symb}(\hat{f})\ast
                {\rm symb}(\hat{g}).
$
The star product is non--commutative, but associative, because
``{\rm symb}''
    provides an algebra homomorphism  between the operator
algebra and the symbols.

In the physical applications we have
in mind, the Hilbert space ${\cal H}$ is the state space of a
quantum mechanical system, and the manifold
${\cal M} \equiv {\cal M}_{2N}$
                                                   is the
 $2N$--dimensional classical phase--space of this system.
 Quantum mechanical operators $\hat{f}$ are represented
 by functions $f = f (\phi)$, where $ \phi^a =
 (p^1, \cdots, p^N, q^1, \cdots, q^N), a = 1, \cdots,
 2 N$ are canonical coordinates on ${\cal M}_{2N}$.
 We assume that the phase--space
 has the topology of ${\bf R}^{2N}$, and that
 the symplectic 2--form
 $\omega = \frac{1}{2} \omega_{ab} d \phi^a \wedge d \phi^b$ has
constant components:
$\omega_{j,N+i}=-\omega_{N+i,j}=\delta _{ij}$.
The inverse of this matrix,
$
\omega^{ab}$,
defines the Poisson bracket
$
\{ f, g \}_{\rm pb} =
\partial_a f \omega^{ab} \partial_b g
$.
Specifying a symbol
map means fixing an ordering prescription, because it
associates a unique operator $\hat{f} (\hat{p}, \hat{q}) =
{\rm symb}^{-1} (f (p, q))$ to
$f (p,q)$. We shall mostly use the Weyl symbol which
associates the Weyl--ordered operator
$\hat{f}$ to any polynomial $f$.
The corresponding star product reads
\begin{equation}
(f \ast g) (\phi) = f (\phi) \exp \left [ i \frac{\hbar}{2}
\stackrel{\leftarrow}{\partial_a} {\omega}^{ab} \stackrel{\rightarrow}
{\partial_b} \right ] g (\phi)
\end{equation}
The commutator with respect to this star product defines the
well-known Moyal bracket:
$
\{ f, g \}_{\rm mb} =
(f \ast g - g \ast f)/i\hbar .
$

Our basic idea is to use the symbol map
to establish a one--to--one correspondence
between the abstract non--commutative differential forms
in $\Omega^p A$ and functions of $p + 1$ arguments,
as well as between the various operations ($\delta$, etc.)
defined on them.
First we extent the notion of a symbol
to the elements of
$A^{\otimes (p+1)}$
in a straightforward way:
\begin{equation}
\left[ {\rm symb} \left( \hat{a}_0 \otimes  \cdots
\otimes \hat{a}_p \right) \right]
\left( \phi_0, \cdots, \phi_p \right)
 = \left[ {\rm symb} \left( \hat{a}_0 \right) \right]
\left(\phi_0 \right)
   \cdots
 \left[ {\rm symb} \left( \hat{a}_p \right) \right]
\left( \phi_p \right)
\end{equation}
As a consequence,
the map $\delta$~has a natural action on the generalized symbols:
\begin{equation}      \delta \,{\rm symb} \left( \alpha_p \right)\  =\
{\rm symb} \left( \delta \alpha_p \right)   \end{equation}           For
$\alpha_p$'s of the type (\ref{3.16}) one can work out the explicit
form of
\begin{equation}
\delta : {\rm Fun} \left({\cal M}_{2N}^{(p+1)}\right)
\longrightarrow {\rm Fun} \left( {\cal M}_{2N}^{(p+2)} \right)
\end{equation}
Interestingly enough, one finds that
$
 \delta F_p
$
is given by eq.
 (\ref{2.2}), on which also our
discussion of the {\it classical} exterior algebra was based.
There remains
a crucial difference however. The forms
$F_p \in \bigwedge_{\rm MS}^p({\cal M}_{2N})$ studied in the
introduction were supposed to vanish when two adjacent arguments
are equal. The symbols $F_p = {\rm symb} \left( \alpha_p \right)$,
instead, obey a deformed version of this condition, namely
${\rm m}_i F_p = 0$. In the classical
limit $\hbar \rightarrow 0$, when the star--product
which is implicit in the multiplication map becomes the
ordinary point--wise product of functions, the two notions
coincide and
the conditions are the same in both cases.
We conclude that in the classical limit
$\Omega^p A$ and $\bigwedge^p_{\rm MS}({\cal M}_{2N})$ are
equivalent:
\begin{equation}
\fbox{$ \displaystyle
\lim_{\hbar \to 0} \, {\rm symb} \, \left( \Omega^p A
\right) = \bigwedge\nolimits^p_{\rm MS} ({\cal M}_{2N})
$}
\end{equation}

There exists also a very natural definition of a Lie derivative
acting on the universal forms and their symbols.
Let us fix a certain $\alpha_p \in \Omega^p A$ with an
expansion of the type (\ref{3.16}) and let us perform the same
unitary transformation (generated by $\hat{H}$) on all factors
of the tensor product. This leads to the ``time'' evolution
\begin{equation}
i \hbar \, \partial_t \, \alpha_p (t)   =  \sum\limits^p_{j=0}
\sum\limits_{m_0 \cdots m_p} \alpha_{m_0 \cdots m_p}\, \,
                             \hat{a}_{m_0} (t) \otimes \cdots
 \otimes \left[ \hat{H}, \hat{a}_{m_j} (t) \right]
\otimes \cdots \otimes \hat{a}_{m_p} (t)
\end{equation}
If we apply the Weyl symbol map to both sides of this
equation
          we arrive at
\begin{equation}
- \partial_t F_p \left( \phi_0, \cdots, \phi_p; t \right) =
{\cal L}^{\hbar}_p \left[ H \right] F_p \left( \phi_0,
\cdots, \phi_p; t \right)
\end{equation}
with the ``quantum deformed Lie derivative''
\begin{equation}
{\cal L}^{\hbar}_p \left[ H \right] =
- \sum\limits^p_{i=0} \, \frac{2}{\hbar} \, H
\left( \phi_i \right) \sin
\left[ \frac{\hbar}{2} \, \stackrel{\leftarrow}{\partial}_a^{(i)}
                                                   \omega^{ab}\,
\stackrel{\rightarrow}{\partial}_b^{(i)} \right]
\end{equation}
where
$H \equiv {\rm symb} \left( \hat{H} \right)$.
In the limit $\hbar \rightarrow 0$, ${\cal L}_p^\hbar$
reduces to the classical Lie derivative for
multiscalars. This suggests the interpretation of the symbols
${\rm symb} (\alpha), \alpha \in \Omega A$, as quantum
deformed multiscalars. When a classical multiscalar
is subject to a canonical transformation, the hamiltonian
vector field $-V_H = \omega^{ba} \partial_a H \partial_b$
acts on any of its arguments. In the non--commutative case,
$V_H$ is replaced by its Moyal analogue.
The quantum Lie derivative
commutes with the differential $\delta$,
and it gives rise to a closed,
$W_{\infty}$-type algebra \cite{winf}:
\begin{equation}
\left[ {\cal L}_p^\hbar \left[ H_1 \right],
{\cal L}_p^\hbar \left[ H_2 \right] \right] =
- {\cal L}_p^\hbar \left[ \{H_1, H_2\}_{\rm mb} \right]
\end{equation}

It is quite instructive to study
the coincidence limit of the Moyal
multiscalars $F_p = {\rm symb} \left( \alpha_p \right)$.
Differences relative to the classical discussion in the introduction
occur because the pointwise product of functions
on ${\cal M}_{2N}$ is now replaced by the star product.
One of the consequences of this deformation is that,
contrary to the classical multiscalars in $\bigwedge^p_{\rm MS}
({\cal M}_{2N})$, the Moyal multiscalars $F_p$ are not simply
proportional to $ \eta_1^{a_1} \cdots \eta_p^{a_p}$ in
the coincidence limit: there are also terms with
$ \eta_1^{a_1} \cdots \eta_l^{a_l}, l < p$.
As an example, let us look at the
2--form              \begin{equation}        \alpha_2 =
\delta   \hat{a}_0 \delta \hat{a}_1
=  \left[ 1 \otimes \hat{a}_0 - \hat{a}_0 \otimes 1 \right]
\otimes_A \left[ 1 \otimes \hat{a}_1 - \hat{a}_1 \otimes 1 \right]
\end{equation}
By applying the symbol map and expanding the arguments we are led
to
\begin{eqnarray}
 F_2 \left( \phi, \phi + \eta_1, \phi + \eta_2 \right)
 & = & \left( a_0 a_1 - a_0 \ast a_1 \right)
 \left( \phi \right) + \eta_1^b \partial_b
 \left( a_0 a_1 - a_0 \ast a_1 \right) \left( \phi \right)
 \nonumber                                   \\
 & & + \eta_1^b \eta_2^c\  \partial_b a_0
 \left( \phi \right) \partial_c a_1 \left( \phi \right) +
 O \left( \eta_1^2, \eta_2^2 \right)
\end{eqnarray}
This corresponds to
an inhomogeneous differential form at the classical level.
The term proportional to
$\eta_1 \eta_2$ is the expected 2--form evaluated on the vectors
$\eta_1$~and $\eta_2$, but there is also
a term linear in $\eta$ (1--form) and
a constant piece (0--form).
For the quantum analogue of the symplectic 2--form,
$
\hat{\omega}_q = \omega_{ab} \, \delta \hat{\phi}^a
       \delta \hat{\phi}^b
$, this entails that its symbol
\begin{equation}
\omega_q \left( \phi_0, \phi_1, \phi_2 \right) =
\omega_{ab}
\left[ \phi_1^a \phi_2^b - \phi_0^a \phi_2^b + \phi_0^a \phi_1^b \right]
+ i N \hbar
\end{equation}
consists of the classical piece (precisely the symplectic area
of the parallelogram with vertices
$\phi_0, \phi_1$ and $\phi_2$) augmented by a purely imaginary
quantum correction
$ i N \hbar$.
Since, for $N = 1$, $\hat{\omega}_q$ is the
volume form, this term can be related to the fact that the
``quantum volume'' is bounded below, i.e.,
that quantum states cannot be localized in a phase-space volume smaller
than
$( 2 \pi \hbar)^N$. For a more detailed discussion of these
issues we have to refer to \cite{mr}.

\end{document}